\documentclass{ws-ijmpb}
\usepackage{bm}

\begin{document}

\markboth{Tao Zhou, Pei-Ling Zhou, Bing-Hong Wang, Zi-Nan Tang, and Jun Liu}
{Modeling Stock Market Based on Genetic Cellular Automata}

\catchline{}{}{}{}{}

\title
{Modeling Stock Market Based on Genetic Cellular Automata}

\author
{Tao Zhou$^{1,2}$,  Pei-Ling Zhou$^{1}$,  Bing-Hong Wang$^{2}$,  Zi-Nan Tang$^{1}$,  and  Jun Liu$^{1}$}

\address
{$^{1}\!$Department of Electronic Science and Technology,\\
University of Science and Technology of China,\\
Heifei Anhui, 230026, PR China\\
$^{2}\!$Department of Modern Physics and The Nonlinear Science Center,\\
University of Science and Technology of China,\\
Heifei Anhui, 230026 PR China}

\maketitle

\begin{abstract}
An artificial stock market is established with the modeling method and ideas
 of cellular automata. Cells are used to represent stockholders, who have the
 capability of self-teaching and are affected by the investing history of the
 neighboring ones. The neighborhood relationship among the stockholders
 is the expanded {\it Von Neumann} relationship, and the interaction among them
 is realized through selection operator and crossover operator. Experiment
 shows that the large events are frequent in the fluctuations of the stock price
 generated by the artificial stock market when compared with a normal process
 and the price returns distribution is a $L\acute{e}vy$ distribution in the central part
followed by an approximately exponential truncation.
\end{abstract}


\section*{}

\noindent Financial markets are typical complex systems in which the large-scale dynamical
 properties depend on the evolution of a large number of nonlinear-coupled subsystems.
 The efficient market hypothesis (EMH) based on rational expectation assumption
 considers the price of financial markets a random walk, thus the variety of price is unpredictable.
In the recent years, the EMH suffers the impugnation on rational expectation
 assumption and the challenge of actual financial data\cite{1}, and some financial markets
 models are established, including behavior-mind model\cite{2,3}, dynamic-games model\cite{4},
multi-agent model\cite{5,6,7,8,9}, and so on.

Cellular automata(CA) model is a special multi-agent model in which the topological structure is fixed.
It is widely applied in both natural science and social science\cite{10,11}. In this paper, an artificial
stock market based on genetic cellular automata is established. Cells are used to represent
 stockholders, who has the capability of self-teaching and are affected by the investing
history of the neighboring ones. The topological structure of CA in this paper is a two-dimensional
square lattice with periodic boundary conditions, which can be considered as the expanded
{\it Von Neumann} relationship. The information flow will not be cut off by the array-edge in this
network, and since the network is vertex-transitive\cite{12}, each individual has complete symmetrical
position in the corresponding network.

Before a trade, each stockholder should choose the trading
strategies: to buy, to sell or to ride the fence. The
stockholder's decision includes two steps: first, each stockholder
works out a preparatory decision according to the history of its
investment and the stock price. The stockholders of different
risk-properties have different decision methods. The risk-neutral
individuals directly inherit the last decision. The risk-aversed
individuals' investing strategy is to buy at a low price and to
sell at a high price. For an arbitrary risk-aversed individual
{\bf A}, if the average price of {\bf A}'s shares in hand is
$\langle p\rangle$ (if {\bf A} haven't any shares in hand, we let
$\langle p\rangle$ be the mean price of the stock) and the present
stock price is $s(t)$, then {\bf A}'s decision is to depend on the
parameter $x=(s(t)-p)/p\in (-1,+\infty )$. She will chose to buy
at the probability $\sqrt{1-(x+1)^2}$ when $x\leq 0$ and to sell
at the probability $\frac{2}{\pi}\arctan 2(2+\sqrt{3})x$ when
$x>0$, otherwise, she will hold shares. The risk-taking
individuals tend to buy at a up-going price and to sell when the
price is down-going. According to the present up-going range of
stock price $c=(s(t)-s(t-1))/s(t-1)\in (-1,+\infty )$, they will
chose to buy at the probability $\frac{2}{\pi}\arctan 10\sqrt{3}x$
when $x\geq 0$, and to sell at the probability
$\frac{2}{\pi}\arctan (-10\sqrt{3}x)$ when $x<0$, otherwise, they
will do nothing. The individuals' risk-properties are given
randomly in initializing process, and can change along with the
evolvement of the stock market. If an individual chose to buy or
to sell, she should determine the price and amount of the
trading-application. The buying-price and selling-price will be
chosen completely randomly in the interval $[s(t),1.1s(t)]$ and
$[0.9s(t),s(t)]$ respectively. The trading-amount is proportional
to the quantity of capital owned by swapper. After that, each
stockholder starts to investigate its neighbors and change its
decision (even its risk-property) at a certain probability. That
is, the next risk-property and the final decision of an individual
is obtained by choosing neighboring individual to carry on genetic
operation. Considering that stockholders are always inclined to
listen to the winners, the individual beneficial coefficient
$\kappa$ is set as the ratio of current capitalization to initial
capitalization. Then the individual's fitness is $F=\kappa ^{h}
(h\geq 0)$, where $h$ is called the influence factor. The gap
between winners' and losers' influence will grow larger if $h$
grows bigger. Among the four neighbors, the central individual
will select one to run the crossover operation at the probability,
between whom and the individual's fitness there's a direct
proportion. The risk-property and the decision of each individual
are naturally divided into four types of genes logically,
including risk-property, deal-decision, price-decision and
amount-decision. In this paper, the four genes' crossing over
operation is independent from each other. The offspring draws its
final decision or changes its risk-property by selecting the genes
of the central individual at the probability of $\lambda$ , and of
the neighboring individual at the probability of $1-\lambda$ . The
parameter $\lambda \in [0,1]$ is called individual independence
degree, which describes the affection of neighboring ones upon the
individual. It should be emphasized that the stockholder's
risk-property and final decision may be changed at a very small
mutation probability. The buyer with higher price and the seller
with lower price will trade preferentially, and the trading-price
is the average of seller's and buyer's price. The stock price is
the weighted average of trading-price according to the
trading-amount.

When proper initial condition and parameters have been chosen, the
artificial stock market can generate its stock price whose trend
and fluctuations are rather similar to that of real stock market.
The figure 1 gives a simulating experimental result. In the
experiment we set the market size as $40\times 40$ (i.e. 1600
stockholders), the initial stock price as 2.30, the total quantity
of shares as 5 million units and the total quantity of fund as 10
million. The initial quantity of fund and shares owned follows
normal distribution, and the mutation probability is 0.02. The
initial risk-properties of individuals are drawn randomly.

\vspace{3mm}

\centerline{\scalebox {0.7} [0.38] {\includegraphics {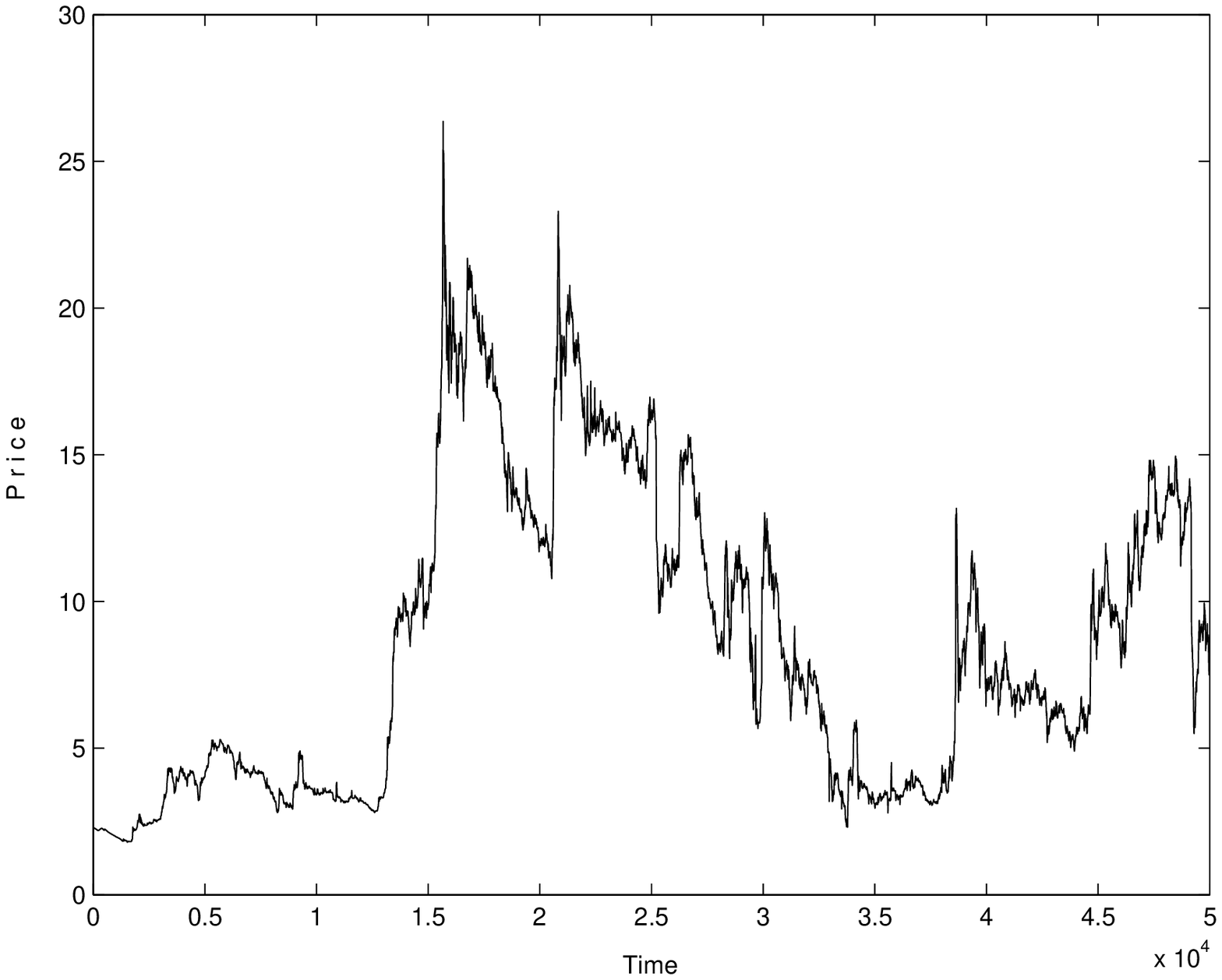}}}
\centerline{{\bf Fig. 1.}  Time series of the typical evolution of the stock price, where $h=1$ and $\lambda =0.4$ }

\vspace{3mm}

Let $\boldsymbol{P}(t)$ denote the stock price time series, the price returns $\boldsymbol{Z}(t)$ are defined as the difference between two
successive logarithms of the price:

\begin{displaymath}
\boldsymbol{P}(t): \boldsymbol{Z}(t)=log\boldsymbol{P}(t+\Delta t)-log\boldsymbol{P}(t).
\end{displaymath}

\noindent The corresponding price returns as $\Delta t=1$ are shown in figure 2.

\vspace{3mm}

\centerline{\scalebox {0.7} [0.3] {\includegraphics {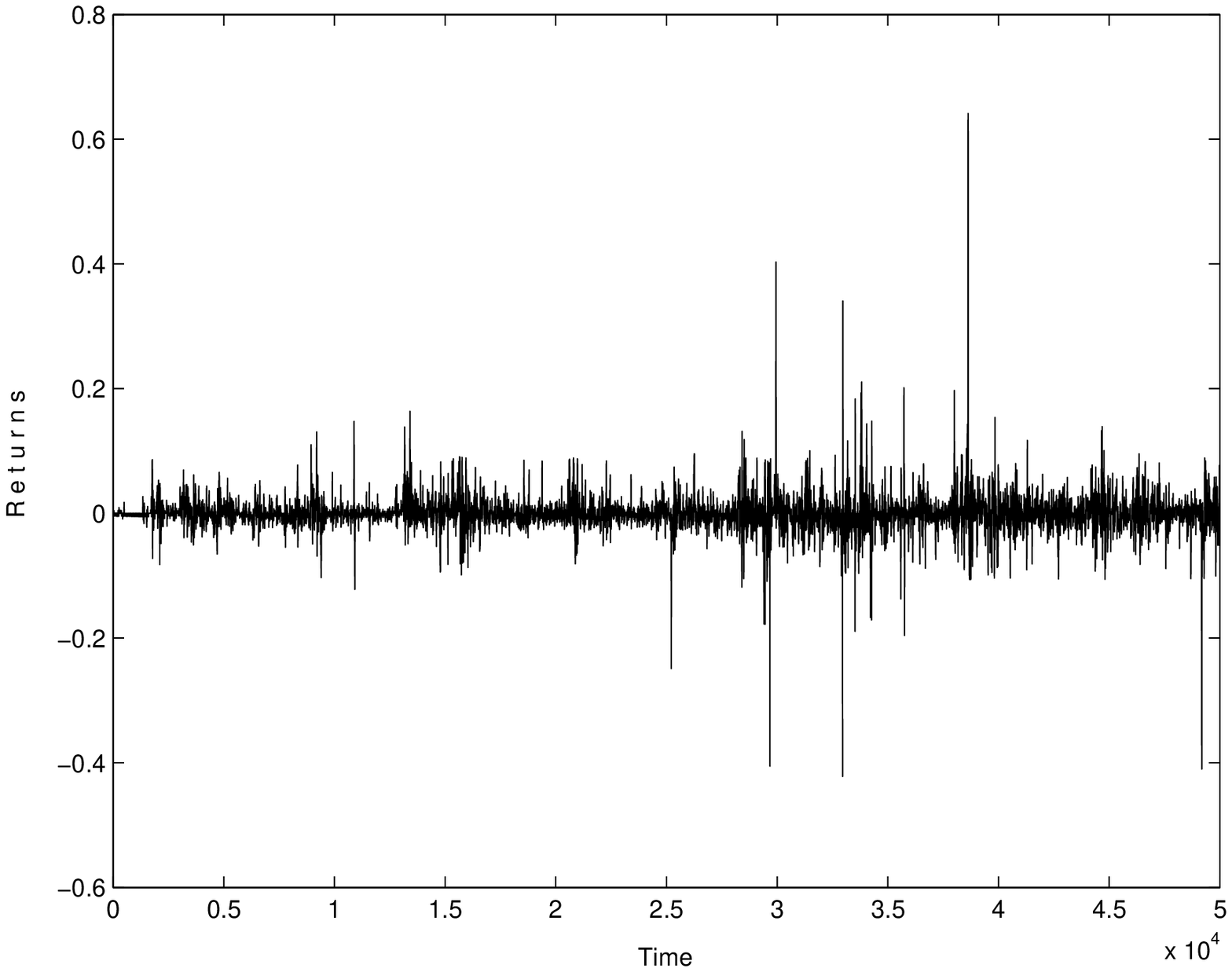}}}
\centerline{{\bf Fig. 2.} The corresponding price returns as $\Delta t=1$}

\vspace{3mm}

{\it Mandelbrot} proposed that the distribution of returns is consistent with a $L\acute{e}vy$ stable distribution\cite{13}.
In 1995, {\it Mantegna} and {\it Stanley} analyzed a large set of data of the S\&P500 index.
It has been reported that the central part of the distribution of S\&P500 returns appears to be well fitted by a $L\acute{e}vy$ distribution,
but the asymptotic behavior of the distribution shows faster decay than that predicted by a $L\acute{e}vy$ distribution\cite{14,15}.
The similar characteristic of the distribution of returns is also found in Heng Seng index\cite{16}.
Figure 3 shows the probability distributions of price returns for $\Delta t$=1,2,4,8,16,32,64.

\vspace{3mm}

\centerline{\scalebox {0.5} [0.5] {\includegraphics {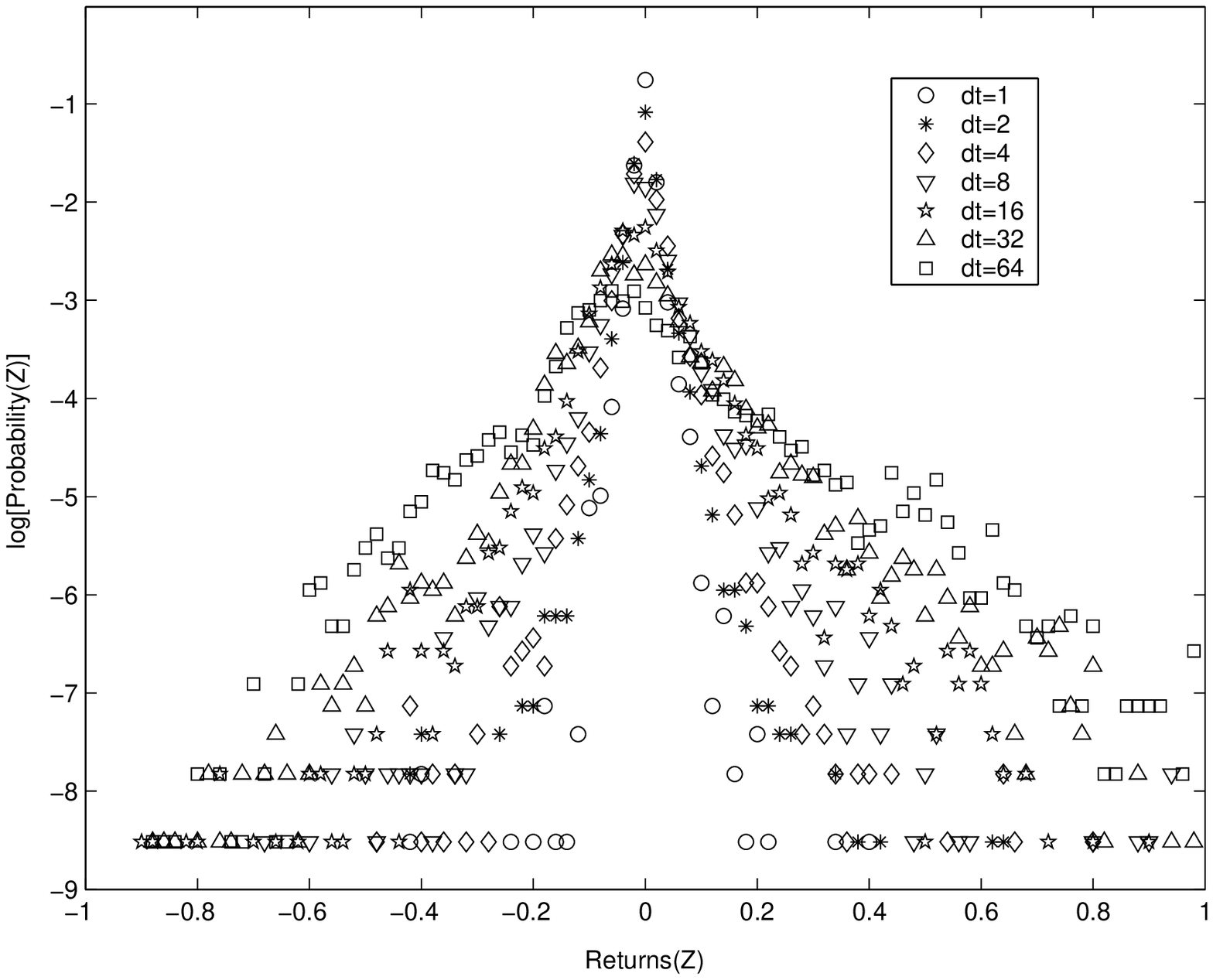}}}
\centerline{{\bf Fig. 3.} The probability distributions of price returns for $\Delta t$=1,2,4,8,16,32,64}

\vspace{3mm}

In figure 2 and figure 3, it can be seen that large events are
frequent in the fluctuations of the stock price generated by the
artificial stock market when compared with a normal process. We
also studied the peak values at the center of the distributions,
figure 4 shows the central peak value versus $\Delta t$ in a
$log-log$ plot. It can be seen that all the data can be well
fitted by a straight line with a slope -0.5632. This observation
agrees with theoretical model leading to a $L\acute{e}vy$
distribution.

In this article, a stock market model is established based on
genetic cellular automata, who has some key characteristics
according with the real-life stock market. Some other experiments
(not include in this paper) indicate that the interaction among
individuals will give rise to clusters and herd behaviors, which
may be the possible mechanism that lead to the existence of large
events in our model. In addition, the mutation is very important
too. Further researches can reveal the multi-level of this system,
the process of the forming and damaging of the self-similar
structure in cellula space, etc. Since the main goal of this
article is to establish and describe the model itself, we won't
give detailed experiment results and analyzing, which will be
given elsewhere. \vspace{3mm}

\centerline{\scalebox {0.4} [0.4] {\includegraphics {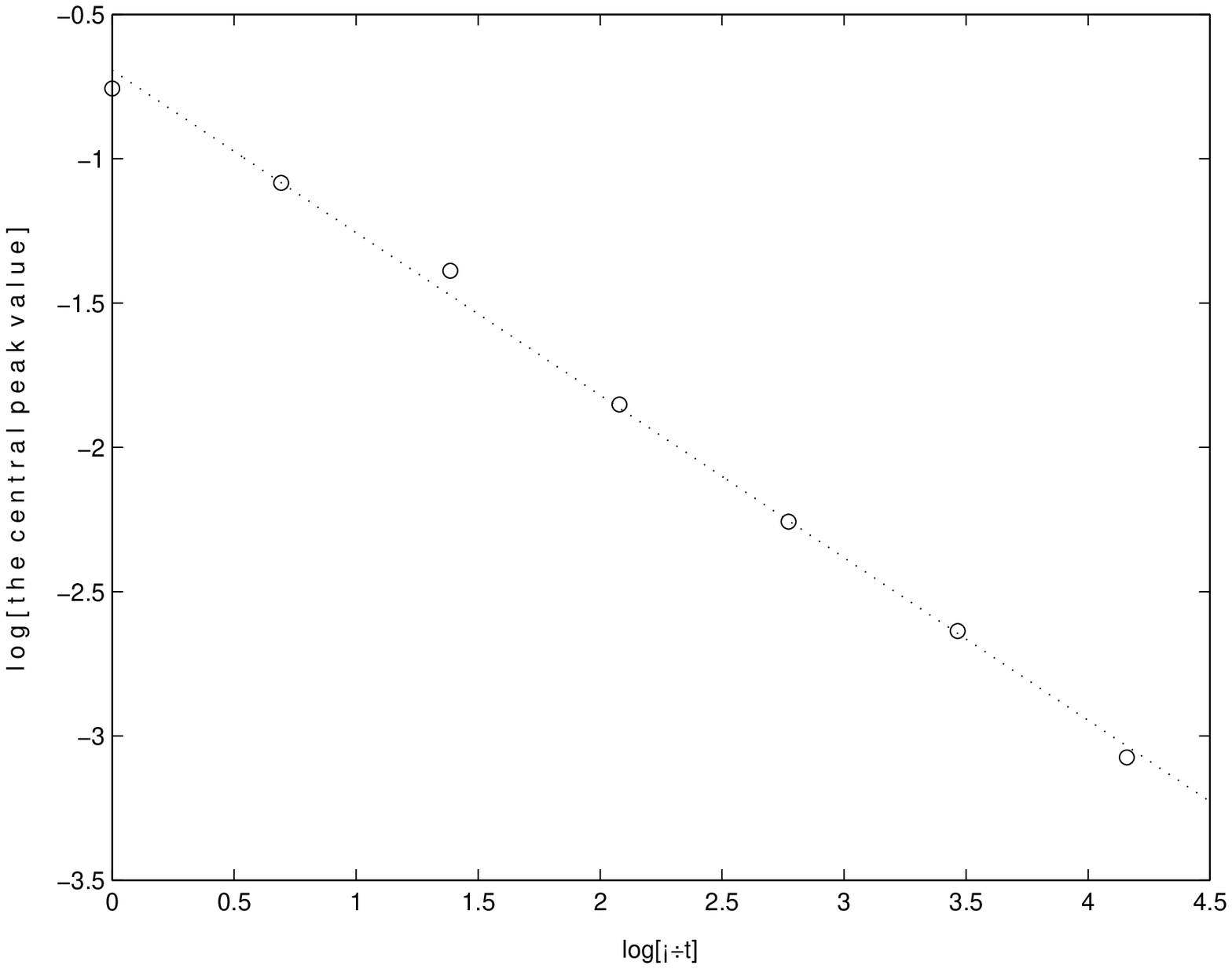}}}
\centerline{{\bf Fig. 4.} The central peak value as a function of $\Delta t$}

\section*{Acknowledgements}

\noindent Supported by the State Key Development Program of Basic Research of China (973 Project), the National
Natural Science Foundation of China under Grant No. 70171053 and 70271070, the China-Canada University
 Industry Partnership Program (CCUIPP-NSFC No.70142005), and the Doctoral Fund from the Ministry of Education of China.

\end{document}